\documentclass[preprint2,12pt]{aastex}
\usepackage{emulateapj5}
\usepackage{onecolfloat5}
\usepackage{natbib}

\newcommand{\mc}{\multicolumn}
\newcommand{\expnt}[2]{\ensuremath{#1 \times 10^{#2}}}   % scientific notation
\newcommand{\gsim}{\stackrel{>}{{\sim}}}
\newcommand{\lsim}{\stackrel{<}{\sim}}

\newcommand{\rxj}{RX~J1856.5$-$3754}
\newcommand{\sgr}{Sgr~dwarf~galaxy}
\newcommand{\hst}{\textit{HST}}

\begin{document}

\shorttitle{The Parallax and Proper Motion of \rxj}
\shortauthors{Kaplan, van~Kerkwijk, \& Anderson}
\twocolumn[
\slugcomment{Accepted by ApJ}

\title{The Parallax and Proper Motion of \rxj\ Revisited}
\author{D.~L.~Kaplan}
\affil{Department of Astronomy, 105-24 California Institute of
Technology, Pasadena, CA 91125, USA}
\email{dlk@astro.caltech.edu}
\author{M.~H.~van~Kerkwijk}
\affil{Sterrenkundig Instituut, Universiteit Utrecht, Postbus 80000,
3508 TA Utrecht, The Netherlands}
\email{M.H.vanKerkwijk@phys.uu.nl}
\and
\author{J.~Anderson}
\affil{Astronomy Department, University of California, Berkeley, CA
94720-3411, USA}
\email{jay@astron.berkeley.edu}

\begin{abstract}
\rxj, a bright soft X-ray source  believed to be the nearest
thermally emitting neutron star, has commanded and continues to command
intense interest from X-ray missions. One of the main goals is to determine
the radius of this neutron star. An integral part of the determination
is an accurate parallax.  \citet{wal01} analyzed
\textit{Hubble Space Telescope} (\hst) data and derived a parallax,
$\pi=16.5\pm 2.3\,$mas.  Combining this distance with the angular radius
derived from blackbody fits to observations of \rxj\ with
\textit{ROSAT}, \textit{EUVE}, \textit{HST}, \citet{pwl+01} 
derived an observed radius (``radiation radius''), $R_{\infty}= 7\,$km.
This value is smaller than the radii calculated from all proposed
equations-of-state (EOS) of dense baryonic matter \citep{hae01}.
Here, we have analyzed the same \hst\ data and find
$\pi=7 \pm 2$~mas.
We have verified our result using a number of different, independent
techniques, and find the result  to be robust.  
The implied radius of \rxj\ is $R_{\infty}=15\pm 6\,$km, falling
squarely in the range of radii, 12--16 km, expected from
calculations of neutron star structure for different equations of
state.  The new distance also implies
a smaller age for \rxj\ of 0.4~Myr, based on its association with the Upper
Sco OB association.  
\end{abstract}

]
\section{Introduction}

The \textit{ROSAT} all-sky survey identified six  neutron stars that
are radio-quiet but bright in
the  soft X-ray band. These sources, unlike the well studied radio
pulsars, lack significant non-thermal emission and are thus excellent
candidates for X-ray spectroscopic studies of the atmospheres of
neutron stars (for reviews, see \citealt{m00,ttzc00}).

The brightest of these sources is \rxj\ \citep*{wwn96}.  A faint, blue
optical counterpart was  identified from \textit{Hubble Space
Telescope} (\textit{HST}) data \citep{wm97}.  \rxj\ has been
intensively studied by most major facilities, especially \textit{ROSAT}, \textit{EUVE},
\textit{ASCA} and \hst. The broad-band data can be well fitted by thermal
emission from a neutron star, which has resulted in the determination of
the effective temperature, $kT_{\rm eff}\approx 50$~eV, and angular radius of the
neutron star, $R_{\infty}/d\approx 0.11\mbox{ km pc}^{-1}$ \citep{pztn96,pwl+01}.

\citet[][hereafter W01]{wal01} used \textit{HST} data spanning three years to measure the
astrometric parameters of \rxj, finding a parallax of $16.5 \pm
2.3$~mas and a proper motion of $332 \pm 1\mbox{ mas yr}^{-1}$ at a
position angle of $100.3\degr \pm 0.1\degr$ ($\mu_{\alpha}=326.7 \pm
0.8\mbox{ mas yr}^{-1}$, $\mu_{\delta}=-59.1 \pm 0.7\mbox{ mas
yr}^{-1}$).

Combining this parallax with the broad-band modeling yields a radiation
radius of $R_{\infty}\approx\!7\,$km.  For the canonical mass of a neutron star,
$1.4\,M_\odot$ this radius is smaller than the
minimum radiation radius of $R_{\infty}^{\rm min}=10.7$~km allowed by
General Relativity \citep{lp00}.
For the radius to exceed $R_{\infty}^{\rm min}$
the mass has to be less than $1\,M_\odot$ \citep{pwl+01}.

The importance of \rxj\ as a laboratory for dense matter physics has
motivated deep observations by \textit{Chandra} \citep{bzn+01},
\textit{XMM} and the Very Large Telescope (VLT). Results
from the first 50-ksec Chandra 
observations\footnotemark\footnotetext{In early October, 2001, 
Chandra observed \rxj\ for an 
additional 450-ksec under the aegis of the Director's discretionary
program.}
can be found in \citet{bzn+01}; the blackbody fits are similar to
those of
\citet{pwl+01}. Using the VLT, \citet{vkk01} discovered an unusual
H$\alpha$ 
nebula around \rxj, from which they infer properties of
\rxj's energetics and emission characteristics \citep{vkk01b}.

Parallax or distance is essential to obtaining the radius, the key
physical parameter (since it now appears that X-ray and optical observations
yield reliable values for $T_{\rm eff}$ and the angular radius).  In view of the
perplexing radius inferred from the parallax measurement of
W01 
we undertook an analysis of the publicly available \hst\ data.
Here we present a detailed description of our analysis followed by
our measurement of the parallax. 

\section{Observations, Analysis \& Results}
\label{sec:ana}

We analyzed the publicly available \hst/WFPC2 observations described by W01 and
\citet{wm97};  see Table~\ref{tab:obs} for a log of the observations.
As noted by W01, the observations took place near the times of
parallactic maximum (3~October and 30~March).  \rxj\ is always on the
Planetary Camera (PC) detector so we only analyzed those data.

\begin{deluxetable}{l c c c c c c c c}
%\rotate
\tabletypesize{\footnotesize}
\tablecaption{WFPC2 Observation Summary\label{tab:obs}}
\tablewidth{0pt}
\tablehead{
\colhead{Epoch} & 
\colhead{$t_{\rm exp}$} & \colhead{$n_{\rm dither}$\tablenotemark{a}} & \colhead{Sky Level} & \colhead{Nominal} &\colhead{$\Delta{\rm PA}$\tablenotemark{b}} &
\colhead{Scale} & \colhead{$x_{0}$\tablenotemark{c}} & \colhead{$y_{0}$\tablenotemark{c}}\\ 
\colhead{(UT)} & \colhead{(s)} &  &
\colhead{$(\mbox{DN s}^{-1})$} & \colhead{PA (deg)} & \colhead{(deg)} & \colhead{(mas pixel$^{-1}$)} & \mc{2}{c}{pixels} \\
}
\startdata
1996~Oct~6 & 4800 & 4 & 0.006&  129.38 & 0.012(2) & $45.5936(8)$ & $408.004(5)$ &$428.143(5)$\\
1999~Mar~30 & 7200 &8 & 0.006& $-$51.75 &$0.0212(7)$ & $45.6028(5)$ & $429.221(3)$ &
$437.824(3)$\\
1999~Sep~16\tablenotemark{d}  & 5191 &4 & 0.005 & 124.97 & 0 & $45.5938$ & 417.948 & 436.803\\
\enddata
\tablenotetext{a}{The dither pattern was in sets of four images, with $(\Delta x$,$\Delta y)=$(0,0), (0,0),
(5.5,5.5), and (5.5,5.5) pixels.} 
\tablenotetext{b}{Defined as the difference between the fitted
position angle  and the nominal position angle (from the image header
and Table~\ref{tab:obs}).}
\tablenotetext{c}{$x$ and $y$ pixel coordinates of the pointing
center, which has $\alpha=-18^{\rm h}56^{\rm m}35\fs374$ and
$\delta=-37\degr54\arcmin31\farcs71$; see \S\ref{sec:absast}.}
\tablenotetext{d}{The values for this epoch were assumed to be correct.}

\tablecomments{See Eqn.~\ref{eqn:tform} for the sense of the transformation.}
\end{deluxetable}

\subsection{Relative Astrometry}
\label{sec:rel}
We used the effective point-spread function (ePSF) fitting technique to
perform the astrometry, as described by \citet[][hereafter
AK00]{ak00}.  We did not have sufficient numbers of stars to derive our
own ePSF for each data set, so we used a previously determined  ePSF (from
archival data) for
the \hst/WFPC2 F555W filter.  While the parallax data were taken with the F606W
filter, we feel that using the F555W ePSF was appropriate, as it was
of superior quality to the F606W ePSF that we have (also derived from
other data).  The difference in
ePSFs should not bias the data, as the wavelength dependence of the ePSF is not strong, especially
across the $\approx 50$~nm difference in effective wavelength between
the filters. Furthermore, the blue
color of \rxj\ brings its ePSF closer to the F555W ePSFs  of normal stars.
In any case, we also 
performed the
analysis with the F606W ePSF. Since the latter ePSFs were of inferior
quality (owing to a less ideal data set), we obtained larger 
errors, but the results were entirely consistent with those 
obtained using the F555W ePSF.

We fit the F555W ePSF to the raw images, uncorrected for dithering or
cosmic rays.  For each epoch, we used a $\chi^{2}$-minimization, as 
described by AK00, to derive a position for each star in
each of the raw images.  We corrected this position for the 34th-row
anomaly\footnote{It is a common error to apply the
34th-row correction \textit{after}  shifting  and combining the data.
This procedure is incorrect, as the 34th-row correction
should be applied to the raw 
image coordinates and not those that have been shifted and
rebinned.} \citep{ak99} and geometric distortions in the PC detector
using new coefficients (Anderson 2002, in preparation).  For each
epoch, this yielded four (or
eight) positions for each star.  

\begin{figure}[hf]
\plotone{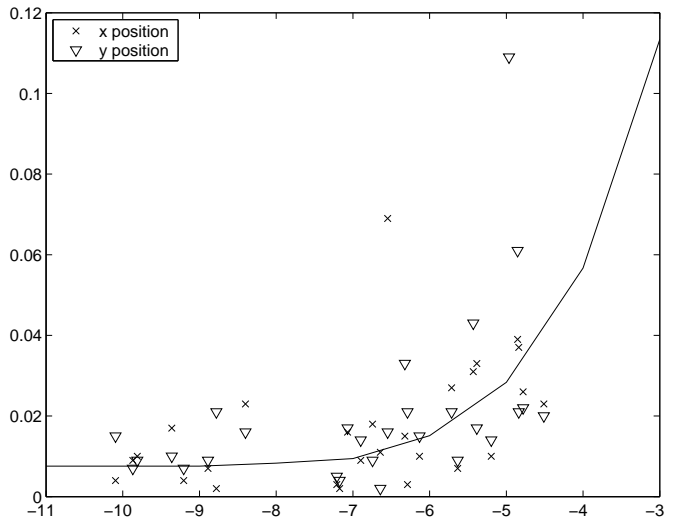}
\caption{1-D position uncertainty (pixels) vs.\ instrumental
magnitude, defined by $m=-2.5 \log_{10}({\rm DN})$ within a $5\times5$
pixel region in a single exposure.  The x's are for the raw
uncertainties in the $x$ positions of the stars used to 
register the different epochs, the triangles for the
raw $y$ uncertainties, and the solid line is a relation determined
from $\sim 5000$ well-observed stars in other data sets.  
The sources here are those that are not saturated and were used in the
analysis; see Tables \ref{tab:ref} and \ref{tab:measxy}.
The data from this paper (x's and triangles) generally follow the trend
defined by the line (also given in Eqn.~\ref{eqn:err}), but there is
considerable spread due to the small 
number of measurements (3 or 4) used to construct each uncertainty.
\label{fig:uncert}}
\end{figure}

We then solved for the shifts (due to dithering)
between the four (or eight) images in a given
epoch; these shifts are given in Table~\ref{tab:dith}.  We rejected sources which had
significantly higher residuals than other sources of their
magnitude (see Figure~\ref{fig:uncert}), ascribable to the source
being extended or saturated. For the remaining sources we used
an iterative $\sigma$-clipping (with threshold at 2.5$\sigma$)
to reject outlier position determinations. Following this
the remaining position measurements were used to derive
the average position for each source.  The final distortion-corrected
source positions in image $(x,y)$ coordinates as well as the
number of accepted measurements are listed in Table~\ref{tab:measxy}.

\begin{deluxetable}{l c c c}
\tablecaption{Fitted Dither Positions\label{tab:dith}}
\tablewidth{0pt}
\tablehead{
\colhead{Epoch} & \colhead{Image} & \colhead{$\Delta x$} &
\colhead{$\Delta y$} \\
 & \colhead{Number} & \mc{2}{c}{(pixels)} \\}
\startdata
1996~Oct~6 &1&	\phs  0.000& \phs 0.000 \\
           &2&	   $-$0.049& $-$0.112 \\
           &3&	\phs  5.451& \phs 5.417 \\
           &4&	\phs  5.416& \phs 5.270 \\\tableline
1999~Mar~30&1&	\phs  0.000& \phs 0.000 \\
     	   &2&	\phs  0.022& \phs 0.115 \\
     	   &3&	\phs  5.456& \phs 5.455 \\
     	   &4&	\phs  5.465& \phs 5.476 \\
     	   &5&	   $-$0.066& $-$0.142 \\
     	   &6&	   $-$0.097& $-$0.064 \\
     	   &7&	\phs  5.368& \phs 5.628 \\
     	   &8&	\phs  5.399& \phs 5.532 \\\tableline
1999~Sep~16&1&	\phs  0.000& \phs 0.000 \\
           &2&	   $-$0.159& $-$0.027 \\
           &3&	\phs  5.515& \phs 5.542 \\
           &4&	\phs  5.386& \phs 5.422 \\

\enddata
\end{deluxetable}

In the limit of a large number of independent
observations that are well dithered, the measured rms would have
yielded reliable uncertainties. However, we are limited by the 
meager number of available frames at each epoch.
With a maximum of four (or even eight) measurements of each position
we were not able to search for systematic errors. Furthermore, 
the data were taken with non-optimal dithering\footnote{For
astrometric purposes, a $2\times2$ dithering grid is minimal for
optimal removal of pixel-phase errors; a 15-point dither pattern is
even better.  See AK00.}, with the fractional pixel portions of the
dither often repeating from one image to the next
(Table~\ref{tab:dith}).  This increases the likelihood of systematic
pixel-phase errors.
We therefore adopted a semi-empirical
approach for the measurement uncertainties along the following lines.

As the first 
approximation of the uncertainty for
each position, we take the rms variation between the positions used to
construct the average.  As the next level, we used the expected precision 
in the positions as a function of the signal-to-noise ratio
(SNR) of stars. 
To this end, we utilized an astrometric database that one of us (J.A.)
has built up over the last several years. In particular, 
we used 
18 well-dithered PC images 
that were obtained under similar conditions (filter, background,
crowding) to those discussed here. In Figure~\ref{fig:uncert}
we display a fit (obtained from the measurements of about 5000 stars)
to the astrometric precision as a function of
the SNR of stars. 
The relation from Figure~\ref{fig:uncert} is reasonably well fit by
\begin{equation}
\sigma = \frac{1}{\sqrt{2}}\left[ \left(2.38 e^{0.69 m}\right)^{3}
+ (0.02)^{3}\right]^{1/3}\mbox{ pixels},
\label{eqn:err}
\end{equation}
where $m = -2.5 \log_{10}({\rm DN})$ (within a $5 \times 5$~pixel
area) is the magnitude measured in a single exposure and $\sigma$ is
the 1-D position uncertainty.  For $N$ well-dithered exposures, the
uncertainty is $\sigma/\sqrt{N}$, as expected (AK00).

We see that the raw uncertainties
generally follow the expected trend, but that there is substantial
scatter.  This is not surprising, given that we may be computing the
uncertainties from four or fewer than four measurements.  Therefore, in our
analysis we use the maximum of the empirically determined uncertainty
for an individual star
and the uncertainty from the relation in Figure~\ref{fig:uncert}
corrected to the appropriate magnitude;
the stars that have uncertainties larger than those inferred from this
relation do so due to cosmic rays or proximity to bright sources.

The above analysis gives us reliable and accurate measurements of
stellar positions, but while these stars are in the background
relative to \rxj\ they can still have their own motions that will bias
our determinations.  Therefore, to have some idea of the absolute motion of the stars in the
image, we included in the data measurements of the positions of two
slightly extended sources (presumed to be galaxies) present on the
\hst\ images (see Table~\ref{tab:ref}).  As these
sources are non-stellar, we could not use the AK00 technique to measure
their positions.  Instead we fitted Gaussian profiles and then
applied the
same distortion corrections as with the other technique.  Gaussian fitting is
inherently less accurate than ePSF fitting (AK00), but the errors are important primarily for sources that are
undersampled by the WFPC2 pixels (i.e.\ where pixel-phase errors are important).  The galaxies were reasonably well
resolved (${\rm FWHM} \approx 3.2$~pixels for source \#20,
$\mbox{FWHM} \approx 3.6$~pixels for source \#104), so they should not
suffer from systematic errors related to undersampling.

\begin{deluxetable}{c c c c c c c c c c c c}
\tabletypesize{\footnotesize}
\tablecaption{Distortion Corrected ($x,y$)  Source Positions\label{tab:measxy}}
\tablewidth{0pt}
\tablehead{
\colhead{ID} & \mc{11}{c}{Epoch} \\
 & \mc{3}{c}{1996~Oct} & \colhead{} &\mc{3}{c}{1999~Mar} &\colhead{}&\mc{3}{c}{1999~Sep} \\  \cline{2-4}
\cline{6-8} \cline{10-12}
 & \colhead{$x$} & \colhead{$y$} & \colhead{$N$\tablenotemark{a}} & \colhead{} &
\colhead{$x$} & \colhead{$y$} & \colhead{$N$\tablenotemark{a}} & \colhead{} &
\colhead{$x$} & \colhead{$y$} & \colhead{$N$\tablenotemark{a}} \\ 
 & \mc{2}{c}{(pixels)} & & &\mc{2}{c}{(pixels)} &&  & \mc{2}{c}{(pixels)}
& \\}
\startdata
100 & $269.63(3)$ & $173.83(2)$ &  3  &  & $562.43(2)$ & $694.92(2)$ &  8  &  & $304.53(2)$ & $170.38(2)$ &  4  \\
102 & $452.45(1)$ & $132.70(1)$ &  4  &  & $379.55(3)$ & $732.33(1)$ &  8  &  & $489.55(2)$ & $146.83(1)$ &  4  \\
103    & $583.81(1)$ & $107.54(1)$ &  4  &  & $247.36(2)$ & $755.08(2)$ &  8  &  & $623.26(1)$ & $134.02(1)$ &  4  \\
104 & $612.51(3)$ & $120.50(5)$ &  4  &  & $218.50(5)$ & $741.63(7)$ &  8  &  & $650.93(5)$ & $149.61(5)$ &  4  \\
105    & $616.08(4)$ & $295.90(4)$ &  4  &  & $218.82(4)$ & $565.99(4)$ &  8  &  & $637.53(4)$ & $324.77(4)$ &  3  \\
106    & $561.03(2)$ & $366.40(2)$ &  4  &  & $275.145(9)$ & $496.71(1)$ &  8  &  & $576.066(8)$ & $389.601(8)$ &  4  \\
107 & $84.80(2)$ & $400.36(2)$ &  4  &  & $751.80(2)$ & $471.926(9)$ &  8  &  & $98.673(9)$ & $378.58(1)$ &  4  \\
108    & $233.92(33)$ & $725.84(11)$ &  4  &  & $609.04(8)$ & $144.08(9)$ &  7  &  & $216.36(12)$ & $716.08(11)$ &  4\\
110    & $278.43(2)$ & $762.35(2)$ &  4  &  & $565.18(2)$ & $106.41(2)$ &  8  &  & $257.51(2)$ & $757.17(2)$ &  4  \\
111    & $738.04(1)$ & $728.20(2)$ &  4  &  & $104.91(1)$ & $131.54(1)$ &  8  &  & $718.50(2)$ & $766.57(1)$ &  4  \\
112    & $707.599(9)$ & $519.420(9)$ &  4  &  & $131.642(9)$ & $340.811(9)$ &  8  &  & $707.46(2)$ & $555.82(2)$ &  4  \\
113    & $507.48(7)$ & $598.59(2)$ &  4  &  & $332.92(2)$ & $265.71(3)$ &  8  &  & $501.25(3)$ & $615.74(5)$ &  3  \\
114    & $523.36(2)$ & $522.78(3)$ &  4  &  & $315.58(1)$ & $341.14(1)$ &  8  &  & $524.10(1)$ & $541.72(2)$ &  3  \\
116 & $69.84(2)$ & $146.13(1)$ &  4  &  & $761.95(2)$ & $726.29(1)$ &  8  &  & $107.72(1)$ & $124.24(2)$ &  1  \\
117 & $434.37(1)$ & $135.98(1)$ &  4  &  & $397.66(3)$ & $729.44(2)$ &  8  &  & $471.62(30)$ & $147.99(30)$ &  3\\
118    & $452.54(1)$ & $254.49(2)$ &  4  &  & $381.66(2)$ & $610.43(2)$ &  8  &  & $478.38(1)$ & $268.20(2)$ &  3  \\
119    & $597.78(3)$ & $375.85(4)$ &  4  &  & $238.33(3)$ & $486.51(3)$ &  8  &  & $612.02(10)$ & $402.70(13)$ &  3\\
127    & $123.33(4)$ & $660.75(6)$ &  4  &  & $718.28(4)$ & $210.87(4)$ &  8  &  & $112.59(3)$ & $641.37(3)$ &  4  \\
128    & $130.64(3)$ & $624.33(2)$ &  4  &  & $710.32(2)$ & $247.15(2)$ &  7  &  & $123.17(2)$ & $605.92(2)$ &  3  \\
129    & $161.07(4)$ & $534.78(3)$ &  4  &  & $677.93(7)$ & $335.92(3)$ &  8  &  & $162.31(6)$ & $519.85(5)$ &  4  \\
201 & $764.454(8)$ & $494.44(2)$ &  4  &  & $74.173(8)$ & $364.652(8)$ &  8  &  & $766.54(1)$ & $536.370(8)$ &  4  \\
J    & $450.297(9)$ & $374.648(8)$ &  4  &  & $385.946(8)$ & $490.481(8)$ &  8  &  & $465.10(1)$ & $387.523(8)$ &  4  \\
19  & $88.005(8)$ & $303.943(9)$ &  4  &  & $746.47(2)$ & $568.192(8)$ &  8  &  & $111.311(9)$ & $282.994(8)$ &  4  \\
20 & $243.62(4)$ & $455.03(3)$ &  4  &  & $593.76(3)$ & $414.32(3)$ &  8  &  & $252.08(4)$ & $447.98(2)$ &  4  \\
21  & $235.02(1)$ & $380.065(9)$ &  4  &  & $601.15(1)$ & $488.972(9)$ &  8  &  & $250.279(9)$ & $372.88(1)$ &  4  \\
23  & $317.345(9)$ & $444.296(9)$ &  4  &  & $520.60(2)$ & $423.45(1)$ &  8  &  & $325.671(9)$ & $444.27(3)$ &  4  \\
25 & $602.610(8)$ & $650.40(2)$ &  4  &  & $239.092(8)$ & $211.774(8)$ &  8  &  & $590.571(8)$ & $676.538(8)$ &  4  \\
26     & $698.28(2)$ & $682.24(1)$ &  4  &  & $144.40(1)$ & $177.714(8)$ &  8  &  & $682.42(1)$ & $717.680(8)$ &  4  \\
28  & $375.731(8)$ & $290.699(8)$ &  4  &  & $458.94(1)$ & $575.797(9)$ &  8  &  & $398.67(2)$ & $297.003(8)$ &  4  \\
X & $357.93(4)$ & $516.84(4)$ &  4  &  & $472.30(4)$ & $333.92(4)$ &  8  &  & $368.42(4)$ & $540.08(4)$ &  4  \\
\enddata
\tablenotetext{a}{The number of independent measurements used to
determine the mean position; see \S\ref{sec:rel}.}
\end{deluxetable}

\subsection{Registration of Epochs}
\label{sec:reg}
To determine the transformation of the background sources (all sources
except for \rxj) between epochs, we proceeded iteratively. 
Our basic input data set was the 27 stellar sources that had consistent
measurements in each epoch plus the two galaxies
(\S\ref{sec:rel}), given in Table~\ref{tab:measxy}.  The faintest of these sources were 
as faint as \rxj, and the brightest were
$\approx 160$ times as bright as \rxj\ (the brightest non-saturated
sources on the WFPC2 images).  First, we
set the fiducial positions of the sources to their
measured positions at epoch  1999.7.  We chose this epoch
 as the effects of  parallax  between
it and epoch 1996.8 are minimized (due to similar parallactic angles) and
the effects of 
proper motion between it and epoch 1999.3 are minimized (due to close
proximity in time), thus yielding the
best matches to the other epochs given the limited information that we
have. We assumed that the position angle, scale, and pointing center
of this fiducial epoch are known.  The pointing center has no impact
on the results, and is simply a dummy parameter.  The position angle
and scale are known to reasonable precision ($<0.1\degr$ for the
position angle, and $<0.1$\% for the scale).  For our nominal values, we
take the values from the \textit{HST} image headers (listed in
Tables~\ref{tab:obs}).  If these values are wrong
at certain levels,
they would introduce errors on those levels into our measurements
(i.e.\ if the fiducial scale were wrong by 0.1\%, our proper motions
would be wrong by the same amount) but these are systematic effects
that are well below the measurement precision of the current data.

For our transformation between the epochs, we assumed a simple
rotation, scale, and offset.  This is valid if the distortion
correction (\S\ref{sec:rel}) removed all nonlinear terms.  The
transformation equation between the measured positions of star $i$ at
epoch $j$, $(x_{i,j},y_{i,j})$, to the celestial position
$(\Delta \alpha_{i,j},\Delta \delta_{i,j})$ is

\begin{eqnarray}
\Delta \alpha_{i,j} &=& -{\rm scale}_{j}\left((x_{i,j}-x_{0,j})\cos {\rm PA}_{j}
-(y_{i,j}-y_{0,j})\sin {\rm PA}_{j}\right) \nonumber \\
\Delta \delta_{i,j} &=& {\rm scale}_{j}\left((x_{i,j}-x_{0,j})\sin {\rm PA}_{j}
+(y_{i,j}-y_{0,j})\cos {\rm PA}_{j}\right), 
\label{eqn:tform}
\end{eqnarray}
where ${\rm scale}_{j}$ is the plate scale (arcsec per pixel),
$(x_{0,j},y_{0,j})$ are the offsets, and ${\rm PA}_{j}$ is the 
position angle  of epoch $j$, all of which are given in Table~\ref{tab:obs}.

We performed a $\chi^{2}$ fit between the fiducial positions and  the
positions at the three measured epochs, varying the 
scale, position angle, and offsets of the other epochs.    This fit gave
relatively large $\chi^{2}$ values, due to proper motion between the epochs.

\begin{figure*}[hf]
\epsscale{1.5}
\plotone{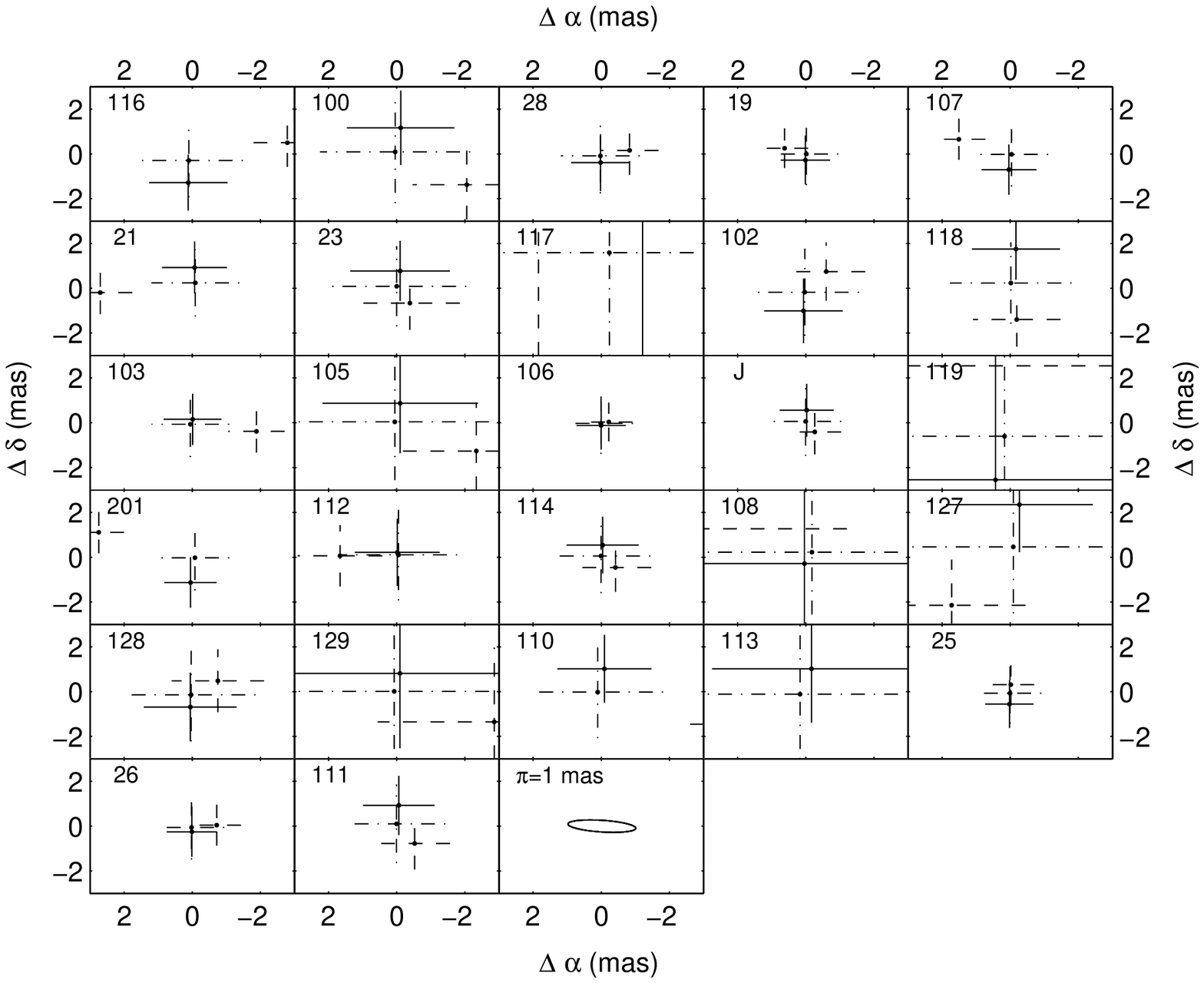}
\caption{Residual positions for background sources with best-fit
proper motion removed;  see Figure~\ref{fig:parellps} for \rxj.
Position from 1996~October has dot-dashed cross; position from
1999~March has dashed cross; position from 1999~September has solid
cross.  All sources are labeled with identifiers from
Table~\ref{tab:ref} in the upper-left corners. The last plot in the
bottom row shows a parallactic ellipse for $\pi=1$~mas.}
\label{fig:stars}
\epsscale{1.0}
\end{figure*}

We then fit for  updated fiducial positions and proper motions of the
background sources based on a linear least-squares technique (for the
galaxies, the proper motion was forced to be 0).  
These positions and proper motions were used to re-calculate the
expected positions in the non-reference epochs, which
dramatically lowered the $\chi^{2}$ values.  We iterated this
procedure (fitting for the  transformation between the epochs, then fitting for the
individual positions and proper motions) making sure that the solution was converging.
After five iterations, $\chi^{2}$ changed by 0.2; we considered the
solution to have converged.  
The results of the fitting (the fiducial positions
and proper motions) are given in Table~\ref{tab:ref}.

For our analysis we did not fit for the parallaxes of the background
sources as they are primarily at distances of $>1$~kpc (see
Appendices~\ref{sec:detl} and \ref{sec:bg}).  Our final $\chi^{2}$
value for the fit was 228 for 50 degrees of freedom.  The  reduced
$\chi^{2}$ deviates significantly from 1, indicating that we may be 
missing some source of error in our analysis.  We have determined,
though, that the majority of the excess $\chi^{2}$ comes from four objects:
the stars \#21 and \#201, and the galaxies \#20 and \#104.  Without these sources, we
obtain a $\chi^{2}$ value of 56 for 38 degrees of freedom, or a
reduced $\chi^{2}$ of 1.5.  This value is much more acceptable.  The
question, then, is  why these four sources contributed so much to the
$\chi^{2}$.  For the galaxies, they were not measured with the ePSF
technique of AK00, and are therefore subject to systematic errors not
addressed here; we only include them to provide an ``inertial'' reference
frame in the analysis.  The stars, \#21 and \#201, are among the brightest of the
sources measured (Table~\ref{tab:ref}) and may be among the closest
(except for \rxj, of course).  As can be seen from their position
residuals (Figure~\ref{fig:stars}) they may have parallaxes of $\sim
1.5$~mas and therefore would not be expected to register well.  They are also close to
extremely bright stars (Figure \ref{fig:pmbg}) that may bias the
position measurements.  Together, these effects (non-zero parallax and
mis-measuring positions) significantly increase the
$\chi^{2}$ for the fit.  Neither of these effects is important
for the majority of the stars, which typically have  residuals
consistent with zero parallax (also see Appendix~\ref{sec:bg}).  For
the results of the registration, the inclusion or exclusion of these
sources does not matter.

\begin{deluxetable}{c c c c r r r r}
\tabletypesize{\small}
\tablecaption{Reference Sources for Astrometry\label{tab:ref}}
\tablewidth{0pt}
\tablehead{
\colhead{ID\tablenotemark{a}} & \colhead{$\alpha$\tablenotemark{b} $-18^{\rm h}56^{\rm m}$} &
\colhead{$\delta$\tablenotemark{b} $+37\degr54\arcmin$}& \colhead{$M_{\rm F606W}$}&
\colhead{$\Delta \alpha$\tablenotemark{c}} & \colhead{$\Delta \delta$\tablenotemark{c}} 
 & \colhead{$\mu_{\alpha}$} & \colhead{$\mu_{\delta}$} \\ 
 & \colhead{(s)} & \colhead{(arcsec)} & \colhead{(mag)}&\mc{2}{c}{(arcsec)} & \mc{2}{c}{(mas yr$^{-1}$)} \\
}
\startdata
100 & 34.28&  $-$29.2 &   25.0&  $-12.9651(9)$ & $  2.4976(9)$ & $0.1(5)$ & $5.6(6)$ \\   
102 & 34.60&  $-$21.6 &   24.5&  $-9.1416(8)$ & $10.0946(9)$ & $-5.6(3)$ & $-6.6(3)$ \\
103 & 34.85&  $-$16.2 &   23.9&  $-6.2193(4)$ & $15.4760(3)$ &$-2.7(2)$ &   $1.9(2)$ \\
104\tablenotemark{d}& 34.96&  $-$15.6 & 25.2 & $-4.924(1)$ & $16.117(1)$ & 0 & 0 \\ 
105 & 35.49&  $-$20.6 &   26.0&  $ 1.3560(2)$ & $11.1550(2)$ & $-1.3(7)$ & $-0.9(7)$ \\
106 & 35.56&  $-$24.5 &   22.4&  $ 2.2431(3)$ & $ 7.1809(3)$ & $-2.7(3)$ &  $0.9(3)$ \\ 
107 & 34.50&  $-$42.3 &   23.7&  $-10.3348(4)$ & $-10.5884(4)$ & $-1.8(3)$ & $-1.2(3)$ \\
108\tablenotemark{e}& 35.83&  $-$46.5 &   25.0&  \nodata & \nodata & \nodata & \nodata\\                    
110 & 36.05&  $-$45.9 &   25.1&  $ 8.0253(7)$ & $-14.2280(7)$&$-0.9(4)$ & $2.4(4)$ \\ 
111 & 37.07&  $-$28.7 &   24.6&  $20.1264(4)$ & $  2.9624(5)$& $1.5(3)$ & $4.8(3)$ \\  
112 & 36.38&  $-$23.8 &   23.6&  $11.8772(5)$ & $ 7.9134(5)$ & $-2.4(2)$ &$-1.0(2)$ \\
113 & 36.12&  $-$33.1 &   23.8&  $ 8.8837(3)$ & $ -1.411(1)$ & $0.1(9)$ & $6(1)$ \\ 
114 & 35.94&  $-$30.4 &   24.2&  $ 6.6715(4)$ & $ 1.3405(4)$ & $-0.3(5)$ & $4.8(4)$ \\ 
116 & 33.71&  $-$35.5 &   24.0&  $-19.7242(6)$ & $ -3.7665(6)$ & $-0.9(3)$ & $-3.3(3)$ \\
117\tablenotemark{e}& 34.57&  $-$22.4 &   25.8&  \nodata & \nodata & \nodata & \nodata\\                   
118 & 34.97&  $-$25.1 &   24.6&  $-4.8365(6)$ & $ 6.5790(6)$ & $-1.9(3)$ & $-4.9(3)$ \\
119\tablenotemark{e}& 35.68&  $-$23.6 &   25.7&  \nodata & \nodata & \nodata & \nodata\\                    
127 & 35.37&  $-$48.5 &   26.3&  $-0.043(1)$ & $-16.758(1)$&$-1(1)$ &    $1.1(9)$ \\
128 & 35.28&  $-$47.2 &   24.9&  $-1.1148(9)$ & $-15.4528(9)$ & $-1.2(5)$ &$-1.2(5)$ \\
129 & 35.09&  $-$43.5 &   26.3&  $-3.374(3)$ & $-11.782(2)$ & $4.5(8)$ & $1.8(9)$ \\ 
201  & 36.44&  $-$21.1 &   20.4&  $12.6471(3)$ & $10.6446(3)$ & $-0.7(2)$ & $0.6(2)$ \\ 
 J  & 35.32&  $-$28.7 &   21.0&  $-0.6627(4)$ & $ 3.0372(4)$ & $-0.4(2)$ & $-0.3(2)$ \\
19  & 34.22&  $-$39.4 &   21.9&  $-13.6277(4)$ & $ -7.6757(4)$ & $2.4(2)$ & $2.4(2)$ \\  
20\tablenotemark{d}& 35.05&  $-$38.3 &  23.7& $-3.8055(8)$ & $-6.5591(8)$ & 0 & 0 \\  
21  & 34.81&  $-$36.4 &   21.0&  $ -6.6871(4)$ & $ -4.7115(4)$ & $3.9(2)$ & $-2.3(2)$ \\ 
23  & 35.20&  $-$35.4 &   23.8&  $ -2.0685(5)$ & $ -3.6799(5)$ & $-5.0(3)$ & $-6.1(2)$ \\
25  & 36.51&  $-$31.3 &   22.0&  $13.4631(2)$ & $  0.4201(2)$& $0.2(3)$ &$-2.6(3)$ \\
26  & 36.84&  $-$28.9 &   21.5&  $17.3585(4)$ & $  2.8446(5)$& $2.2(2)$&$-11.1(3)$ \\
28  & 34.89&  $-$28.9 &   21.7&  $ -5.7783(5)$ & $  2.8325(5)$ & $-0.4(2)$ & $-2.3(2)$ \\ 
 X\tablenotemark{f}& 35.60&  $-$36.2 &   26.1  & $2.643(2)$ & $-4.504(2)$ &\nodata & \nodata \\
\enddata
\tablenotetext{a}{ID's are as in W01 where possible; source 201
was not present in W01; X is \rxj.}
\tablenotetext{b}{Positions are measured at equinox J2000,
epoch 1999.26.}
\tablenotetext{c}{
Position offsets at epoch 1999.7, relative to the
pointing center, for which our absolute astrometry yields
$\alpha=-18^{\rm h}56^{\rm m}35\fs374$ and
$\delta=-37\degr54\arcmin31\farcs71$; see \S\ref{sec:absast}.}
\tablenotetext{d}{Extended source, probably a galaxy.}
\tablenotetext{e}{Rejected from the analysis due to poor position
measurements; see Table~\ref{tab:measxy}.}
\tablenotetext{f}{For proper motions, see Table~\ref{tab:par}.}
\end{deluxetable}

We performed
additional analyses to determine how robust our measurements are, and
included information from these analyses in the final estimates (see
\S\ref{sec:par} and Appendix~\ref{sec:detl}).

The deviations of the  scales and position angles from the nominal
value were small but significant (see Table~\ref{tab:obs}), unlike
stated by W01.  We find
that the scale changed by $\approx 0.03$\% from one epoch to another,
and that the position angle changed by $\approx0\fdg02$.  This is reasonable, given the
fluctuations seen in other WFPC2 observations
(due to thermal fluctuations in the detector and telescope assembly; AK00).

\subsection{Absolute Astrometry}
\label{sec:absast}
Absolute astrometry was done relative to the USNO-A2.0 catalog
\citep{m98}.  We first determined centroids for all 571 USNO-A2.0
stars that overlapped with the average R-band image obtained in 2000
using FORS2 on UT\#2 (Kueyen) at the Very Large Telescope (see
\citealt{vkk01}; the image is composed of 29 exposures of 135\,s).  We
rejected 63 objects that were badly overexposed or had widths
inconsistent with them being stellar.  Next, the pixel coordinates
were corrected for instrumental distortion using a cubic radial
distortion function provided to us by T.~Szeifert and W.~Seifert
(1999, private communication).  Finally, the zero point position, the
plate scale, and the position angle on the sky were determined,
rejecting iteratively a further 87 objects for which the residuals to
the solution were larger than $0\farcs6$ (inspection of the images
showed that virtually all of these were visual doubles, which are
blended on the sky survey plates on which the USNO-A2.0 coordinates
are based).  For the 421 stars that pass our cuts, the inferred
single-star measurement errors are $0\farcs18$ in each coordinate,
which is line with the uncertainties expected for the USNO-A2.0
catalogue \citep{d99}.  Thus, we conclude that our astrometry is tied
to the USNO-A2.0 system at about $0\farcs01$ accuracy.

We used the solution to determine the positions in the VLT R-band
image of 19 stars from Table~\ref{tab:ref} (for the remaining 7 --
objects 105, 106, 112, 113, 118, 127, and 128 -- it was not possible
to determine accurate positions, either because they were too faint or
because they were too close to brighter stars).  Using these epoch
2000.3 positions and the fiducial epoch 1999.7 positions derived from
the registration of the {\em HST} images above, we derived the
pointing center for our reference {\em HST} image (we corrected for
the difference in epoch using the observed proper motions).  We solved
for zero point offsets only, i.e., the scale and orientation were held
fixed to the values listed in the header.  The inferred single-star
measurement errors are $0\farcs025$ in each coordinate, and the
zero-point should thus be tied to the R-band image to better than
$0\farcs01$.\footnote{Leaving the position angle and scale free, we
find changes of $0\fdg09$ and 0.07\%, respectively.  The corresponding
change in inferred pointing center is $\sim\!0\farcs002$.}  With this
pointing center, we determined the absolute positions for all stars
listed in Table~\ref{tab:ref}.  These should be on the USNO-A2.0
system to about $0\farcs02$, and on the International Celestial Reference
System to about $0\farcs2$.

\subsection{Determination of Parallax and Proper Motion of \rxj}
\label{sec:par}
With the three epochs registered, we compared the positions of \rxj\
in each (see Table~\ref{tab:measxy}).  We combined the initial estimates
of the position uncertainties in quadrature with the uncertainties introduced by the
registration.
We fit for the proper motion and
parallax of \rxj\ using a linear least-squares solution.  The
locations along the parallactic ellipse at each epoch were determined
using the JPL DE200 ephemeris.

As noted in \S\ref{sec:rel} and \S\ref{sec:reg}, our limited
number of measurements means that the individual position
uncertainties have limited accuracy.  Because of this, the uncertainty on
the parallax  derived from strict statistical considerations 
(1.7~mas)  may not be correct.
We have therefore estimated the parallax uncertainty using a variety of
techniques (see
Appendix~\ref{sec:detl}); these techniques have an rms of 0.4~mas,
which we add in quadrature to find an rms of 1.8~mas.  
To be conservative we round this up, finding the overall 1-$\sigma$
uncertainty to be 2~mas, similar to the value found by W01.

We present the results of the fitting for $\pi$, $\mu_{\alpha}$, and
$\mu_{\delta}$ in Table~\ref{tab:par}.  We also present the values for
the derived parameters of distance $D$ and transverse velocity
$V_{\perp}$.   The best-fit parallax  is $7 \pm 2$~mas.  We
can exclude a null-result for the parallax at the $10^{-4}$ level.
Our results are inconsistent with  those of W01 at the $99.8$\% level.
However, our best-fit values for the proper motions are entirely
consistent with those presented in W01, and also agree with the
orientation of the H$\alpha$ nebula \citep{vkk01b}.

\begin{deluxetable}{l c}
\tablecaption{Motion of \rxj\label{tab:par}}
\tablewidth{0pt}
\tablehead{
\colhead{Parameter} & \colhead{Best-fit Values} \\
}
\startdata
$\alpha$\tablenotemark{a} & \phs$2.6435 \pm 0.0042$\\
$\delta$\tablenotemark{a} & $-4.5050 \pm 0.0030$\\
$\mu_{\alpha}$ ($\mbox{mas yr}^{-1}$) & \phs$328 \pm 1$ \\
$\mu_{\delta}$ ($\mbox{mas yr}^{-1}$) & \phn$-58 \pm 1$\\
$\pi$ (mas) & $7 \pm 2$ \\[0.25cm]
%\tableline
$D$ (pc) & $140 \pm 40$ \\
$\mu$ ($\mbox{mas yr}^{-1}$) & $333 \pm 1$ \\
PA (deg) & $100.2 \pm 0.2$ \\
$V_{\perp}$ ($\mbox{km s}^{-1}$) & $220 \pm 60$ \\
\enddata
\tablenotetext{a}{Fiducial positions at epoch 1999.7, relative to
$(x_{0},y_{0})$ offsets given in Table~\ref{tab:obs}.}
\tablecomments{Best-fit values for $\mu_{\alpha}$, $\mu_{\delta}$,
$\pi$ are determined directly from a linear least-squares solution, and errors are
1-$\sigma$/68\% confidence.  Best-fit values and errors for the other
parameters are derived from those for $\mu_{\alpha}$, $\mu_{\delta}$ and
$\pi$.}
\end{deluxetable}

The severe inconsistency between the values of the parallax derived by W01
and that derived by us is puzzling.  
The most obvious explanation for the discrepancy between our measured
parallax and that of W01 is the significant difference in the way the data were 
processed: we used an ePSF that accounts for pixel phase errors and
fit the data without manipulations such as rebinning or resampling.
We also use more accurate distortion corrections,
 account for small changes in the scale and position angle of the
observations from their nominal values, and  account for the proper
motion of the background objects.  W01, on the other hand, first
resampled the data, then shifted, then 
rebinned the data, and measured the positions with a PSF that is
independent of pixel phase.  Such analysis is liable to introduce even more pixel
phase errors than those that were originally present. 
However, even if we follow the general method of W01 (resampling and
rebinning, Gaussian fitting, old distortion corrections) we cannot
reproduce a parallax of $16.5$~mas (see  Appendix~\ref{sec:detl}).  

The surest way to resolve the differences between our analysis and
that of W01 will be with the 2001~March \hst data (not yet publicly
available).  This should allow further refinement of the proper motion
and a more robust measurement of the parallax, with a final uncertainty of
$\approx 1.5$--2~mas.

\begin{figure}[hf]
\plotone{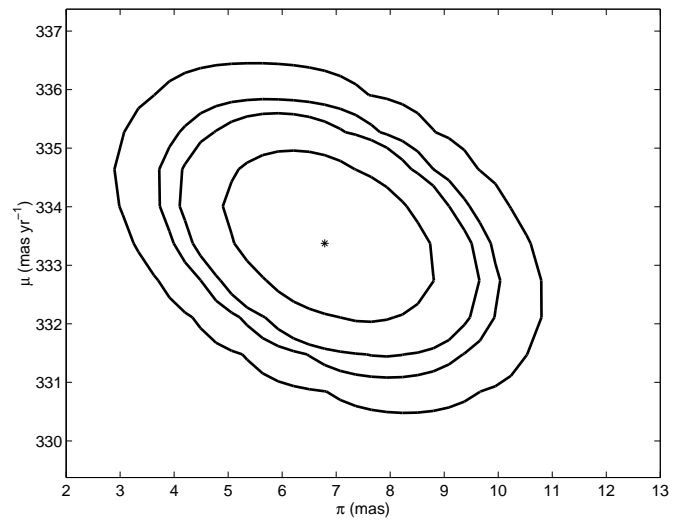}
\caption{Joint confidence contours for the parallax $\pi$ (mas) and the
magnitude of the proper motion $\mu$ ($\mbox{mas yr}^{-1}$) for \rxj,
illustrating the covariance between these parameters.
Plotted are 68\%, 90\%, 95\%, and 99\% confidence contours, with the
best-fit values indicated by the star.  Note that the  contours are for the joint
confidence --- the 1-D 68\%confidence levels are given in
Table~\ref{tab:par}. Compare to Fig.~2 from W01.
\label{fig:cont}
}
\end{figure}

As noted by W01, due to the small angle between the
proper motion and the major axis of the parallactic ellipse, there is
significant anti-correlation between the parallax and the magnitude of
the proper motion.   This is shown  in
Figure \ref{fig:cont}.  We stress, though, that even with the latitude
given by this anti-correlation we cannot accommodate a parallax of 16.5~mas.

\begin{figure}[htf]
\plotone{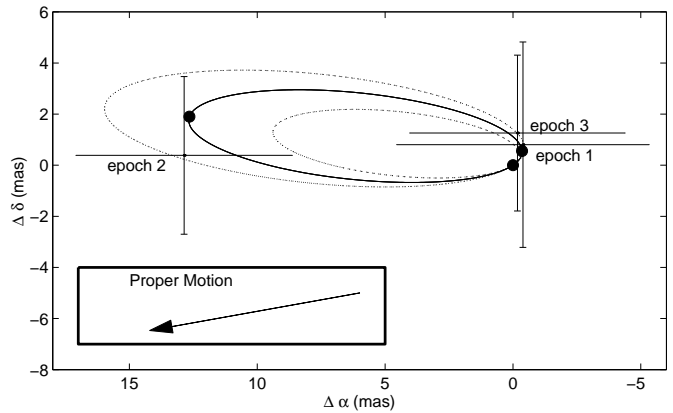}
\caption{Parallactic ellipse for \rxj, showing the measured positions
(points with error-bars) and expected positions (filled circles) at
each epoch (after subtracting the best-fit proper motion).  The inset
shows an arrow that indicates the direction of the proper motion.  The solid ellipse
is for the best-fit parallax of $7$~mas, while the dotted ellipses are
for the $\pm 1$-$\sigma$ values of 5 and 9~mas.  
This figure can be compared to Fig.~4 from W01.
\label{fig:parellps}}
\end{figure}

The position offsets with the proper motion subtracted are shown in
Figure~\ref{fig:parellps}.  The offsets are consistent with the
best-fit parallax.  The correlation between motion due to parallax and
due to proper motion is also illustrated in Figure~\ref{fig:parellps} ---
the proper motion direction differs from the position angle of the
parallactic ellipse by only $20\degr$.

\section{Discussion}
\label{sec:disc}

\subsection{Mass \& Radius}
The most immediate impact of a revised distance for \rxj\ is in the
interpretation of its spectrum.  \citet{pwl+01} used spectral fits to
X-ray and broad-band data to determine a best-fit mass and radius,
taking the previously published distance of $d=61$~pc (W01) to convert
the measured angular size $R_{\infty}/d$ to a radius $R_{\infty}$,
where $R_{\infty}$ is the ``radiation'' radius determined from 
blackbody fitting.  The best-fit radius for all atmosphere choices was $R_{\infty}=6$~km;
causality then requires that the mass be less than this (in geometric units), giving $M < 1
M_{\sun}$ \citep{hae01}.  These values
are inconsistent with all neutron star equations-of-state (EOS), most
of which have radii $R \geq 10$~km \citep{tc99,lp00}.  

Our new distance pushes \rxj\ further away, and therefore allows for
larger radii and masses.  
For example, the best-fit uniform
temperature blackbody fit (a uniform temperature is preferred by the
limits on X-ray pulsations; \citealt{bzn+01}\citealt*{rgs01}) has $R_{\infty}/d=0.11 \pm
0.01 \mbox{ km pc}^{-1}$ \citep{pwl+01}, which had implied  $R_{\infty}\approx
6.7$~km for a distance of 61~pc.  Using our 
revised parallax value changes the radius to $R_{\infty}\approx
15$~km, giving a physical
radius of $R\approx 12$~km (for the canonical 
neutron star mass of $1.4M_{\sun}$; \citealt{tc99}).  This is much more in line with
the likely values for $R$ determined by equations of state
\citep[e.g.,][]{lp00}. 

\citet{p01} has predicted that the passage of \rxj\ near star~\#115
(from W01) in 2003 will cause the apparent position of star~\#115 to
change by $\approx 0.6$~mas due to gravitational lensing.  \citet{p01}
goes on to suggest that precise
measurement of this deflection, perhaps by the new Advanced Camera for
Surveys (ACS) aboard \hst, could be used to measure the mass of \rxj\
and provide an independent estimate to constrain the equation of
state.  
While we do note include star~\#115 in our analysis because it is fainter than
our detection threshold, we were able  to estimate its proper motion.
As expected from its faintness, the measurement is not very accurate;
we find $\mu_{\alpha}=1.2 \pm 1.6 \mbox{ mas yr}^{-1}$, $\mu_{\delta} = 3.4
\pm 1.6\mbox{ mas yr}^{-1}$.  The closest approach of \rxj\ to
star~\#115 is $0\farcs2$ and should occur  around 
April 2004.  With the revised 
distance to \rxj, we find that the Einstein ring radius will be
$\varphi_{\rm E}\approx 9$~mas \citep[eq.\ 2 of][]{p01}, giving a
displacement of $\approx 0.4$~mas.  This is somewhat less than the
displacement predicted by \citet{p01}.  Based on the current WFPC2 data, it appears
{impossible} to measure (in a reasonable number of orbits)
the position of the star with enough precision  to perform the
suggested measurements (even assuming a displacement of 0.6~mas)  with the ACS.  

\subsection{Origin \& Age}

W01 used the observed proper motion of \rxj\ to trace it back towards the
Upper Sco OB association, a source of supernovae during the last few
million years \citep*{dgdzl89}.  He found, for $d=61$~pc, that \rxj\
came within 16~pc of the association 0.9~Myr ago.  W01 notes that for
an unconstrained distance, the radius of closest approach 
is minimized for a distance $d=130$~pc, entirely consistent with
our measurement.  This then gives a closest-approach 
0.4~Myr ago (with a radial velocity of $+30\mbox{ km s}^{-1}$ instead
of $-60\mbox{ km s}^{-1}$), making \rxj\ half the age given in W01.
This revised age  resolves the minor discrepancy in the cooling
history of \rxj\ found by \citet*{ykg01}.

\subsection{Energetics \& Nature}
Another area where the distance enters is in modeling of the
H$\alpha$ nebula that surrounds \rxj\ \citep{vkk01b}.  In most
of the modeling, the distance enters linearly and the factor of $\sim
2$ difference that we find here will not significantly change the
conclusions.  However, there are a number of quantities that have
steeper dependencies on the distance.  We examine each of these.

In their estimate of the minimum pulsar wind
energy loss $\dot{E}$ in the bow-shock model, \citet{vkk01b} find that
$\dot{E} \propto d^{3}$.  We are able therefore to revise the limit to $\dot{E} \gsim
\expnt{8}{32}d_{140}^{3}\mbox{ erg s}^{-1}$, where the distance to
\rxj\ is $140 d_{140}$~pc.  This  impacts on the estimates
of the inferred spin period $P$ and magnetic field
$B$, giving\footnote{Eqns.~8 \& 9 from \citet{vkk01b} contain a small error: both $P$ and $B$
should go as $d^{-3/2}$, not $d^{3/2}$ as written.} $P
\lsim 1.5$~s and $B\lsim \expnt{1}{13}$~G.  The new radial velocity is
reasonably consistent with the inclination angle of $60\degr \pm 15\degr$
determined for the H$\alpha$ nebula.

If \rxj\ were powered by accretion, \citet{vkk01b} find an accretion
rate $\dot{M} = \expnt{3}{9}d_{60}^{-3.5}\mbox{ g s}^{-1}$, where the
distance $d=60 d_{60}$~pc.  For $d_{60}=2.3$, as we find here, this
then implies an available accretion power of $\sim \expnt{3}{28}\mbox{
g s}^{-1}$.  This is now almost four orders of magnitude less than the
observed bolometric luminosity of \rxj, which is revised upwards to 
$\sim \expnt{2}{32}d_{140}^{2}\mbox{ erg s}^{-1}$, further supporting the claim that
accretion cannot power the source.

The final model for the H$\alpha$ nebula considered by \citet{vkk01b}
is an ionization nebula.  Here, \rxj\ can still be a pulsar, but its
energy loss $\dot{E}$ must be small enough that any bow-shock nebula
is smaller than the observed nebula.  This leads to the result 
$\dot{E}\lsim \expnt{2}{34}d_{140}^{3.5}\mbox{ erg s}^{-1}$, a much less
constraining value than that given in \citet{vkk01b}.  However, we
note that the ionization model has become less likely. With a distance
of 140~pc the shape of the ionization nebula cannot be easily
reproduced \citep{vkk01b}.

\subsection{Local Density of Neutron Stars}
Once the emission characteristics and size of one neutron star are
well determined, they can be
used to calibrate other sources.  As an example, we derive a relation
between the optical magnitude, X-ray blackbody temperature, and distance for
isolated neutron stars, and apply it to the two brightest isolated
neutron stars and PSR~B0656+14, a nearby radio pulsar.

The optical emission from \rxj\ and another isolated neutron star,
RX~J0720.4$-$3125 \citep{hmb+97,kvk98}, is 
very well described by the Rayleigh-Jeans tail of a blackbody,
although at a level slightly above that inferred from blackbody fits 
to the X-ray data \citep{vkk01}.  In particular, $f_{\nu} \propto \nu^{2}$ in the
optical regime.  If we assume that the surface compositions of
isolated neutron stars are similar, then $f_{\nu} \propto R^{2}
kT \nu^{2}/d^{2}$, where $R$ is the neutron star radius,
$kT$ is the effective temperature of the surface, and $d$ is the
distance.  
If the neutron stars all have the same radii, 
we can write
\begin{eqnarray}
d_{100} &=& 1.4 \sqrt{\frac{kT}{57\mbox{ eV}}} 10^{(V-25.7)\;{\rm
mag}/5} \nonumber \\
 & =& 1.4 \sqrt{\frac{kT}{57\mbox{ eV}}} 10^{(B-25.3)\;{\rm
mag}/5},
\label{eqn:d}
\end{eqnarray}
where $d=100 d_{100}\mbox{ pc}$, and $B$ and $V$ are the observed optical
magnitudes. 
Here we use the best-fit blackbody temperature; while this
is not always the best-fit overall to the X-ray emission
\citep{pwl+01}, it is a simple model and the dependence of $d_{100}$
on $kT$ is rather weak, so it suffices.
 We list the implied distances for the three other neutron stars
that have thermally-dominated X-ray emission and optical counterparts
in Table~\ref{tab:dist}.  As a general result of our new parallax, we
decrease the local density of isolated neutron stars by a factor of
$\sim 10$ with respect to that inferred from W01.

\begin{deluxetable}{l c c c c}
\tablecaption{Implied Distances to Neutron Stars\label{tab:dist}}
\tablewidth{0pt}
\tablehead{
\colhead{Name} & \colhead{kT} & \colhead{$V$} & \colhead{$d_{100}$}  &
\colhead{Refs.}\\
 & \colhead{(eV)} & \colhead{(mag)} & \\
}
\startdata
\rxj & 57 & 25.7 & 1.4 & 1,2, this work \\[0.25cm]
%\tableline
RX~J0720.4$-$3125 & 79 & 26.6\tablenotemark{a} & 3.0 & 3,4 \\
RX~J1308.8+2127 & 118 & 28.3 & 6.7 & 5,6 \\
PSR~B0656+14\tablenotemark{b} & 73 & 27.3 & 3.3 & 7,8,9 \\
\enddata
\tablerefs{1 -- \citealt{pwl+01}; 2 -- \citealt{vkk01}; 3 --
\citealt{hmb+97}; 4 -- \citealt{kvk98}; 
5 -- \citealt{shs+99}; 6 -- Kaplan et al.\ 2002, in preparation;
7 -- \citealt*{pwc97}; 8 -- \citealt{kpz+01}; 9 -- Zavlin et al.\
2001, in preparation}
\tablenotetext{a}{$B$ magnitude.}
\tablenotetext{b}{We have taken the temperature of the dominant
blackbody component, and extrapolated the $V$ magnitude from the
Rayleigh-Jeans tail found in the UV.}
\tablecomments{Calibrated using the parallax of \rxj; see
\S\ref{sec:disc}, Eqn.~\ref{eqn:d}.}
\end{deluxetable}

For the radio pulsar B0656+14 we get a distance of
$d_{100} \approx 3.3$, near the low end of the values estimated through
other means (250--800~pc; \citealt*{mdlc00}) but still plausible.
We understand that W.~Brisken will soon have a  VLBA
measurement of the parallax, which should provide independent
confirmation of our results (W.~Brisken 2001, personal communication). 

\acknowledgements
Data are  based on observations with the NASA/ESA Hubble
Space Telescope, obtained from the data Archive at the Space Telescope
Science Institute, which is operated by the Association of
Universities for Research in Astronomy, Inc., under NASA contract NAS 5-26555. 
D.L.K.\ holds a fellowship from  the Fannie and John Hertz
Foundation, and his research is supported by  NSF and NASA. M.H.v.K.\
is supported by a fellowship from the Royal Netherlands
Academy of Arts and Sciences.  J.A.\ acknowledges support from HST grant GO-8153.
We thank S.~Kulkarni and D.~Frail for
valuable discussions.

\appendix
\section{Details of Analysis}
\label{sec:detl}

To test the robustness of our analysis, we performed the same general analysis
but with variations on the input data set and analysis method.  These
variations included combinations of:
\begin{itemize}
	\item Using a six-parameter linear transformation between the
	epochs (instead of the standard four-parameter transformation
	involving a shift, scale, and rotation).
	\item Rejecting the stars with the largest position uncertainties.
	\item Rejecting the stars whose derived proper motions had the largest
	uncertainties.
	\item Rejecting stars \#21 and \#201, and the galaxies \#20
	and \#104.
	\item Rejecting the stars more than 300~pixels ($15\arcsec$) from
	\rxj.
	\item Rejecting up to 10 stars at random from the sample.
	\item Using the F606W ePSF instead of the F555W ePSF (see \S\ref{sec:rel})
\end{itemize}
All of these analyses gave entirely consistent results with rms variance of $0.4$~mas, showing that
our parallax measurements are not biased by any particular
data points.  Comparison of these  parallax determinations
allows us to better estimate the uncertainty in the parallax.  To the
formal error determined from the least-squares fit (1.7~mas), we add (in
quadrature) the 0.4~mas rms found above.

As another test, we used the same algorithm to measure the parallaxes
of all of the other stellar sources in the \hst\ images.  As expected
from their photometric distances (Appendix~\ref{sec:bg}),
there were very few convincing parallax measurements for these sources
(Figure~\ref{fig:stars}).  The
measured parallaxes had a mean of $-0.3$~mas and a variance of
$1.2$~mas.  The variance in the measured parallax was roughly
independent of the brightness of the star, down to the brightness of
\rxj, and is reasonably consistent with our estimation of the uncertainty  of
the parallax of \rxj.  We therefore believe that a conservative
estimate of the 1-$\sigma$ parallax uncertainty to be 2~mas.

Finally, we performed the same analysis but with the initial
astrometry done using more conventional Gaussian fitting and with
older WFPC2 distortion coefficients \citep{hhc+95,tvwm95}, like W01.  
Again, the results were consistent with those found using the more
accurate ePSF fitting.

\section{Background Sources}
\label{sec:bg}
\begin{figure}[ht]
\plotone{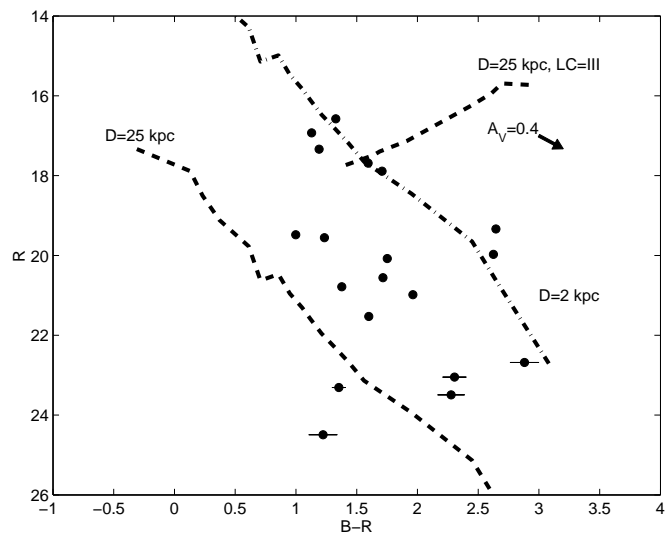}
\caption{Color-magnitude diagram for background
sources. $R$ vs.\ $B-R$, with a $A_{V}=0.4$~mag reddening vector
and a model main sequences at a distance of 2~kpc (dash-dotted line),
a main sequence at a 
distance of 25~kpc (dotted line), and a giant branch at a distance of
25~kpc (dotted line, marked ``LC=III'') from \citet[][p.\ 388, p.\
392]{allen}.  
\label{fig:distbg}
}
\end{figure}

Figure~\ref{fig:distbg} shows the color-magnitude diagram for the
background sources that have reliable VLT photometry \citep{vkk01,vkk01b} and the
distances implied from main-sequence fitting.  Almost all of the
sources are bounded by main sequences with distances from 2--25~kpc,
with a number at implied distances of 10--15~kpc.  Alternately, a
number of the sources are consistent with red-giant branch stars at a
distance $\sim 25$~kpc.

The implied distances of some of these sources, up to 25~kpc, places
then at the edge of the Milky Way and $\approx 2.5$~kpc above the
disk, making them unlikely to be part of the Milky Way.
The sources may be, however, in the \sgr.
The heliocentric distance of the
\sgr\ is $\approx 25$~kpc, and it has a
line-of-sight extent of $\lsim 8$~kpc \citep{iwg+97}. The region near \rxj\ is
$\approx 7.5\degr$ from the center of the \sgr\ \citep{iwg+97},
plausibly within the solid angle subtended by the \sgr.  Therefore a
number of the stars in the field could be main-sequence or giant stars in the \sgr.  

\begin{figure}
\plotone{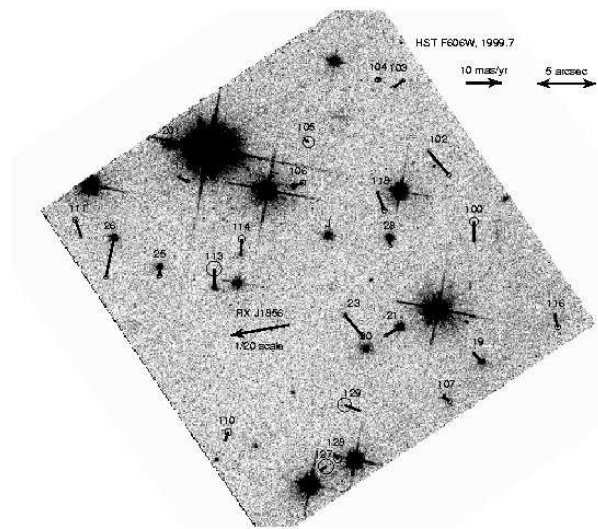}
\caption{Derived proper motions for the background stars and galaxies
used to register the three epochs, overlayed on the 1999.7 \hst\
image.  
The lines indicate proper motions going from the stars to the
circles.  The sizes of the circles indicate the errors in the proper
motions.  The arrow in the upper
right indicates proper motion with a magnitude of $10\mbox{ mas
yr}^{-1}$; next to it is a $5\arcsec$ scale bar.  North is up, and
East is to the left.
The arrow from the position of \rxj\ indicates the
direction but only $1/20$ the magnitude of the source's proper motion (no uncertainties are plotted
for \rxj\ as its proper motion is scaled such that the uncertainties
would be invisible).  All sources are labeled with identifiers from Table~\ref{tab:ref}.
\label{fig:pmbg}}
\end{figure}

In Figure~\ref{fig:pmbg}, we show the proper motion vectors for the
background sources determined from the fitting.  Most of the motions
are small, $<5\mbox{ mas yr}^{-1}$, with the majority being $\approx
2\mbox{ mas yr}^{-1}$.  This is consistent with the magnitude of the
proper motion of the \sgr, $250\mbox{ km s}^{-1}$ at a distance of
25~kpc.  The two galaxies (which were 
forced to have zero proper motion in our 
analysis)  provide an  absolute reference for
these proper motions.  The net proper motion of the background sources
is $\lsim 0.5 \mbox{ mas yr}^{-1}$.  As this is less than the
uncertainty in the measured proper motion for \rxj, the motion of the
background sources should not bias the parallax of \rxj.

\bibliographystyle{apj}
%\bibliography{myrefs,magrefs,casA,xray}

\end{document}